\RequirePackage{fix-cm}

\documentclass{article} 
\usepackage[a4paper,includeheadfoot,margin=2.54cm]{geometry}
\usepackage{graphicx}

\usepackage[
pdfauthor={ },
pdftitle={Predicting interactions between individuals with structural and dynamical information},
pdfsubject={ },
pdfkeywords={ },
pdfproducer={ },
pdfcreator={ }]{hyperref}
 \usepackage{url}      

\usepackage{latexsym}
\usepackage{multirow}

\usepackage{amssymb}
\begin{document}

\title{Predicting interactions between individuals with structural and dynamical information
}


\author{Thibaud Arnoux  --  Lionel Tabourier    --   Matthieu Latapy \\
	\textit{Sorbonne Universit\'{e}, CNRS, Laboratoire d'Informatique de Paris 6, LIP6, F-75005} 	\\
	4 place Jussieu\\ 75005 
    Paris, France\\
    http://www.complexnetworks.fr \\
    \texttt{firstname.lastname@lip6.fr}    
}
\date{ }

\maketitle

\begin{abstract}
	
	Capturing both the structural and temporal aspects of interactions is crucial for many real world datasets like contact between individuals. 
	Using the link stream formalism to capture the dynamic of the systems, we tackle the issue of activity prediction in link streams, that is to say predicting the number of links occurring during a given period of time and we present a protocol that takes advantage of the temporal and structural information contained in the link stream. 
	Using a supervised learning method, we are able to model the dynamic of our system to improve the prediction. 
	We investigate the behavior of our algorithm and crucial elements affecting the prediction. 
	By introducing different categories of pair of nodes, we are able to improve the quality as well as increase the diversity of our prediction.
	\\
	
\textit{Link stream -- Interaction prediction -- Link prediction -- Complex Systems}
\end{abstract}

\section{Introduction}

The study of interaction networks is an important research field with numerous application fields such as security, mobility or recommendation systems.   
They have been studied as graphs for a long time. However, the importance of dynamics has become more and more obvious~\cite{holme2012temporal}. 
Therefore, other representations have been developed to better capture the temporal evolution of these systems. 
We use here the link stream~\cite{DBLP:journals/corr/abs-1710-04073} representation because it fully captures both the structure and the dynamics of interactions.

\begin{figure}
	\begin{center}

		\includegraphics[width=10cm]{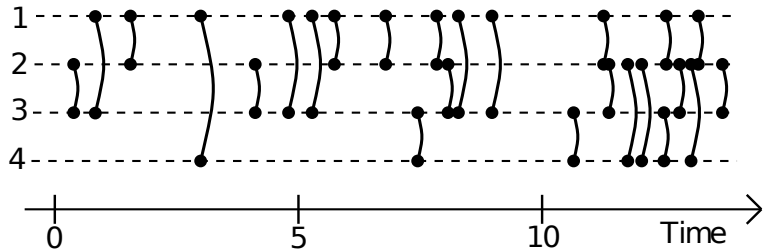}
		\caption{Example of a link stream where nodes $2$ and $3$ have interacted at time $0.5$, nodes $1$ and $3$ have interacted at time $1$, and so on.}
		\label{flot}
	\end{center}
\end{figure}

A link stream (see Figure \ref{flot}) is a sequence of triplets $(t,uv)$, each indicating that an interaction occurred between $u$ and $v$ at time $t$. 
Many real world datasets can be modeled and analyzed using link streams, such as e-mail exchanges, contacts between individuals, phone calls or IP traffic~\cite{DBLP:journals/corr/abs-1710-04073,viard2014identifying,viard2018discovering}.

This modeling is similar to temporal networks \cite{holme2012temporal} or time varying graphs \cite{casteigts2012time}, which hold the same information as link streams. 

	We study here the activity prediction problem, {\em i.e.} predicting the number of interactions that will appear between each pair of nodes during a given period of time.
	While this problem shares properties with the more usual link prediction problem, it is also quite different in the sense that we aim at predicting not only who interacts with who, but also when.
	To do so, we capture independently some structural and dynamical features of link streams properties. 
	Then, we combine these metrics in order to estimate future activity. We present the supervised learning algorithm to optimize the metric combination, and we compare our prediction to ground truth in order to assess the relevance of the approach.
	This shows that combining different types of metrics improves prediction.
	Observing that this approach favors activity prediction of a specific kind of links, we introduce classes of nodes with different behaviors in order to improve the prediction on node pairs.
	We study the performance of our framework on four datasets of real world interactions between individuals~\cite{mastrandrea2015contact,cambridge-haggle-20090529,mit-reality-20050701,roma-taxi-20140717}.
	Finally, we also investigate how different metrics can be used to selectively predict different categories of links and how the introduction of classes allows to preserve diversity in the predicted links while still improving the prediction.	
	Let us emphasize the fact that our goal is to define a general framework to allow further study of the interplay between structural and dynamical features for prediction tasks, rather than optimize our prediction on these specific datasets. 
	Our work provides a general scheme to serve as a baseline and motivation for further work in this direction.
	
		\section{Related work}
	\label{related}

	\begin{figure}[h]
		\begin{center}
		\includegraphics[width=6cm]{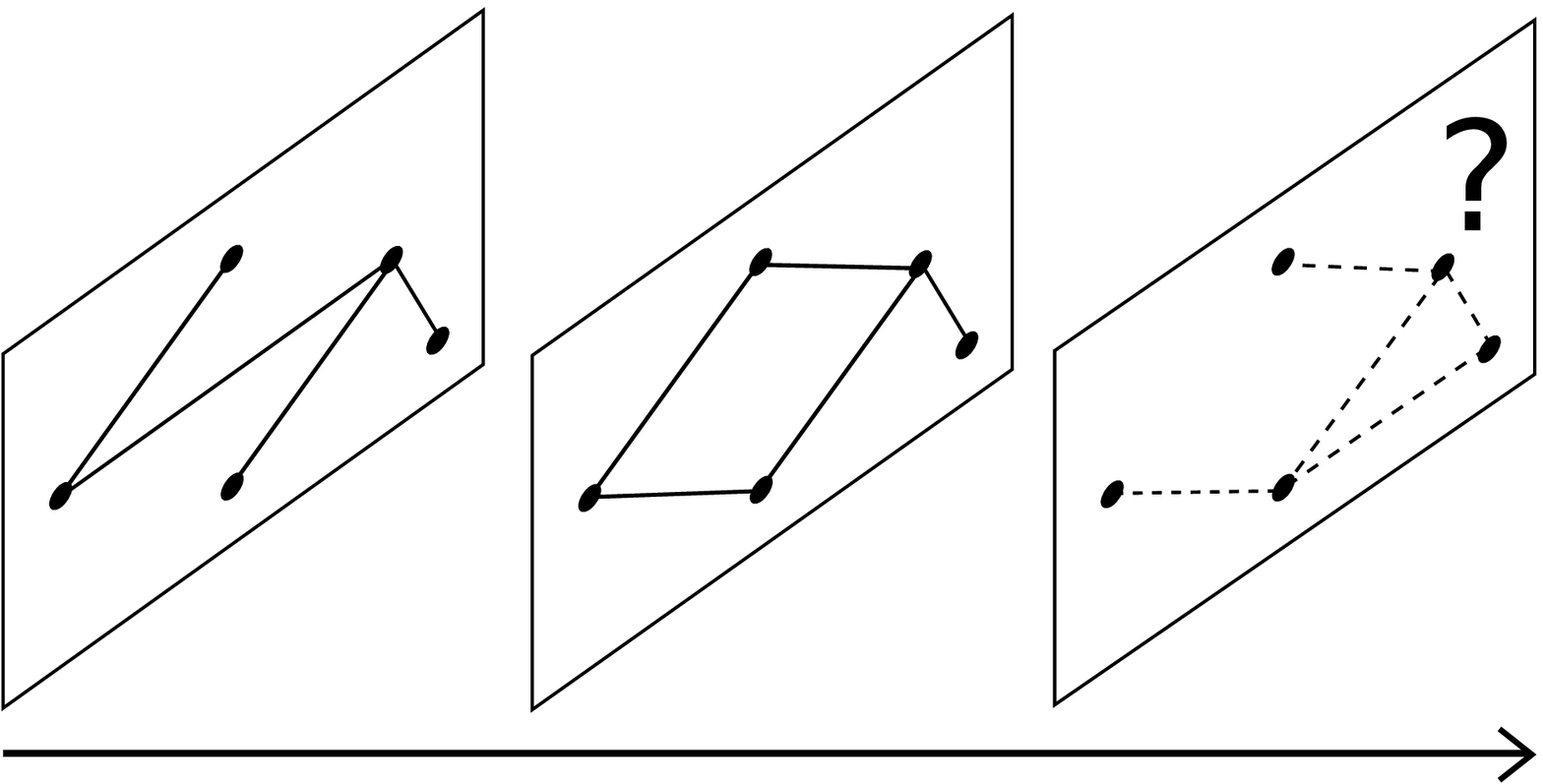}	
		\includegraphics[width=6cm]{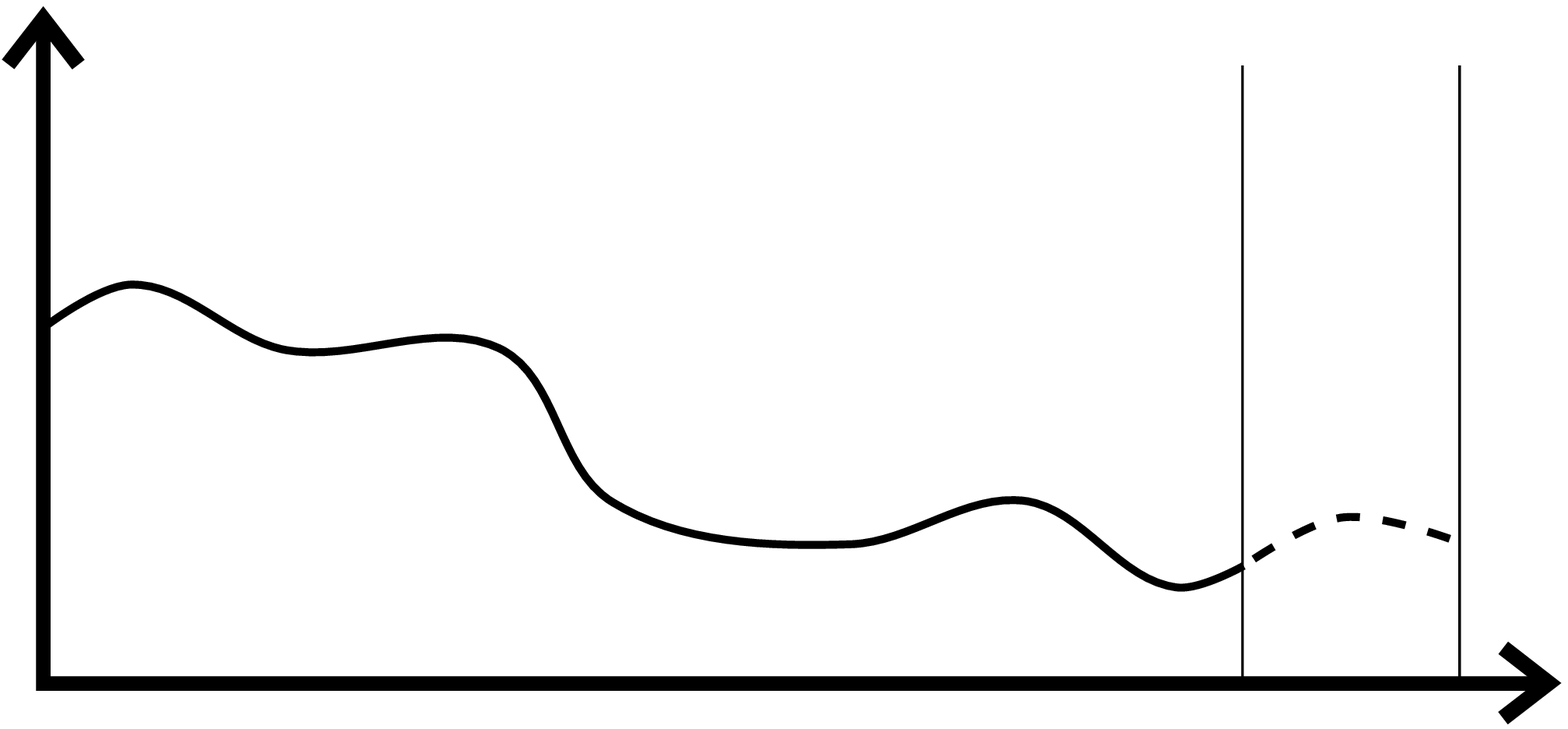}	
		\caption{Basic principle of prediction in dynamic graphs (left) and in times series 
			(right)}
		\label{illustrLPTS}
		\end{center}
	\end{figure}

	Activity prediction is at the crossroad of two classical problems in the literature, namely the link prediction and time-series prediction illustrated in Figure~\ref{illustrLPTS}.
	We briefly present both topics and give more details about works related to our approach.
	Then, we review works using categories in order to improve prediction quality, as we use this technique in our work.

	\subsubsection{Link prediction in graphs}

	In the link prediction problem, the data is represented as a graph \cite{liben2007link,huang2005link,wang2014link}, and the task consists in predicting links to appear in this graph.
	When the temporal evolution is a key element in the data, a usual approach is to slice data into several time windows $T_i$, then aggregate them as a sequence of graphs $G_i=(V,E_i)$ with $E_i=\{(u,v) :  \exists (t,u,v) \in E,~t\in T_i\}$.  
	It allows to use traditional link prediction methods on these graphs.
	The information contained in the data is then extracted using graph-based measurements.
	In this field, many metrics have been developed to obtain the most relevant information~\cite{al2006link,lu2011link}. 
	They often consist in evaluating the similarity between two nodes according to various criteria, which produces a score or a ranking correlated to the apparition probability of a link between these nodes. 
	For example, the number of common neighbors between two nodes and several variants~\cite{kossinets2006effects} are commonly used.
	Similarity measures based on the temporal patterns of activities of nodes and links have also been proposed (\textit{e.g.}~\cite{tabourier2016predicting}).

	Several methods exist to combine the metrics computed for improving the prediction. 
	It is possible to use classification algorithms to determine the predicted links~\cite{al2006link,davis2013supervised}. 
	Another approach is to rank node pairs using the values of different metrics. 
	The predicted links connect the $n$ first node pairs, with $n$ fixed as a parameter and determined using the system behavior \cite{pujari2012supervised,lichtenwalter2010new}. 
	However, the use of time windows commands a time scale and leads to the loss of some temporal information. 
	For example, the information associated to a link repetition between two nodes within a time window disappears. 
	One of the stakes of our work is to conserve this information by using the link stream formalism, better suited to the data.

	Another approach to link prediction using time varying graphs is to aggregate the temporal information in the system by attributing weights to links, based on previous interactions~\cite{murata2007link,tabourier2014rankmerging,dunlavy2011temporal}.
	This process allows to gather structural information as well as indirectly using temporal information in the data.

	\subsubsection{Times series to predict repeated interactions}

	It is also possible to approach link prediction by focusing on the dynamics of interactions between two nodes rather than on the structural properties~\cite{da2012time}. 
	The sequence of links between each pair of nodes is then considered as a time series and numerous tools have been developed to predict the future behavior of such series. 
	For example it is possible to focus on the link apparition frequency in the past to predict future interactions~\cite{tylenda2009towards}. 
	This approach focuses on predicting future occurrences of links that have appeared in the past.
	It is also its main limitation in regards to activity prediction, as it is not suited to predict new links appearing in the system.
	As such, it is complementary to link prediction in graphs. 

	\subsubsection{Link categories}

	To improve prediction quality in a supervised learning problem, it is often efficient to divide items into categories~\cite{lichtenwalter2010new,brockwell2013time}.
	In the case of link prediction, node pairs with similar properties can be gathered in order to model their behavior more efficiently.

	For example it is possible to group node pairs by communities: pair inside communities are more likely to have similar behavior than those going from a community to another~\cite{clauset2008hierarchical}.

	Some study have adopted mixed approaches~\cite{huang2009time}, using both link prediction in graph and a time series approach. However these studies have mostly been focusing on time windows to model the temporal evolution. 
	Our work differs from these methods by focusing on both the dynamic and the structural aspects of the data while avoiding information loss induced by the use of time windows. 
	We introduce a protocol that combines these information sources in a consistent way. We use this to predict both new and repeated links in the stream.

	\section{Problem definition}		
		
	\label{problemdef}

	We consider a set of nodes $V$ representing entities in the system.
	We observe interactions between these entities for a period of time $T=[A,\Omega]$, that we model as a link stream \cite{DBLP:journals/corr/abs-1710-04073} $L=(T,V,E)$, where
	$E \subseteq T\times V\otimes V$, and $(t,uv)\in E$ means that an interaction occurred between $u$ and $v$ at time $t$.
	In the following, we refer to $L$ and $T$ as the \textit{input stream} and \textit{input period}, respectively.
	Our goal is to predict the number of interactions between nodes in $V$ during another period of time $T'=[A',\Omega']$ with $\Omega \leq A' < \Omega'$. 
	We  model the interactions during this interval as a link stream $L'=(T',V,E')$ with $E' \subseteq T'\times V \otimes V$. 
	$L'$ and $T'$ are then called the \textit{prediction stream} and \textit{prediction period}.
	Our aim is to predict the activity of any pair of nodes in the system, \textit{i.e.} for each $(uv) \in V \otimes V$, the value $\mathcal{A}'(u,v)=|\{(t,uv)  \in E'\}|$. Notice that on may take $A' > \Omega$, leading to a prediction period that starts in the future. 

	\section{Activity prediction framework}

	Our approach relies on metrics that capture different features, which a may \textit{a priori} be important for activity prediction and that we describe first. 
	Then, we explain how to combine them into a prediction index, in order to use the different kinds of information captured by our metrics.
	Next, we define our prediction protocol, which consists in first extrapolating a global number of links from the prediction period, then allocating these links to node pairs using the prediction index.
	Finally, we present an evaluation protocol the fits the specificities of our approach. 

	\subsection{Prediction metrics}
\label{metrics}

	The information contained in a link stream can be of different kinds, for instance, it can be the number of past interactions between two nodes or the density of a node's neighborhood.
	In general, existing methods focus either on structural features (in the case of link prediction) or on temporal features (for time-series prediction), while we intend at using metrics adapted to link streams, which combine temporal and structural information.

	\subsubsection{Structural metrics} 
\label{structmetrics}

	As a first step, we adapt metrics from link prediction in graphs to the context of activity prediction in link streams.
	We use traditional metrics widely employed in link prediction, e.g. the number of common neighbors between two nodes $u$ and $v$, and other derived metrics. 
	In order to define them, first we define the neighborhood of a node $u$ in $L$, as $\mathcal{N}(u)=\{v : \exists (t,uv) \in E\}$.
	These metrics are independent of time. 
	This leads to the definition of the following metrics:

	\begin{itemize}

		\item the \textit{Common Neighbors index (CN)} is the number of common neighbors of two nodes:
		
		\[CN(u,v)=|\mathcal{N}(u) \cap \mathcal{N}(v)|\]
		
		\item  the \textit{Jaccard index (JI)} \cite{jaccard1901etude}, which is close to the number of common neighbors, but lays stress on the overlap between the neighborhoods of the nodes:
		
		\[ JI(u,v)=\frac{|\mathcal{N}(u) \cap \mathcal{N}(v)|}{|\mathcal{N}(u) \cup \mathcal{N}(v)|}\]
		
		\item the \textit{Sorensen index (SI)} \cite{sorensen1948method}, which can be seen as a variant of the \textit{Jaccard index}:
		
		\[SI(u,v)=\frac{2 \cdot | \mathcal{N}(u) \cap \mathcal{N}(v)|}{|\mathcal{N}(u)|+|\mathcal{N}(v)|} \]
		
		\item the \textit{Adamic-Adar index~\cite{adamic2003friends} (AA)}, which is especially designed for link prediction, by decreasing the weight of shared neighbors with a high-degree:
		
		\[ AA(u,v)=\sum_{w \in \mathcal{N}(u) \cap \mathcal{N}(v)} \frac{1}{\log |\mathcal{N}(w)|}\]
		
		\item the \textit{Resource Allocation index (RA)} \cite{zhou2009predicting}, close to the \textit{Adamic-Adar index}, but giving weight differently to the degree of shared neighbors:
		
		\[RA(u,v)=\sum_{w \in \mathcal{N}(u) \cap \mathcal{N}(v)} \frac{1}{ |\mathcal{N}(w)|}\]
		
	\end{itemize}

	\subsubsection{Temporal metrics} 
\label{tempmetrics}

	To contrast with graphs, the link stream formalism also captures temporal information.
	As it is usually done in the field of time-series prediction, we first use as a benchmark the extrapolation of past activity, captures link repetitions in the stream.
	To keep it as simple as possible, we use the number of interactions $\mathcal{A}_{(u,v)}$ between $u$ and $v$ occurring during $T$. In the following we refer to this metric as the Pair Activity Extrapolation defined as $PAE(u,v)=|\{(t,uv) \in E\}|$.

	Then, we define two other metrics that describe more precisely the temporal behavior of the system.
	They are adapted versions of the pair activity extrapolation that focus on the most recent activity during the input period.
	This choice is made on the ground that recent interactions affect more the dynamics than the older do~\cite{dunlavy2011temporal}.
	First, we only take into account the activity during a recent period of time of duration $\delta$: for each pair of nodes, we compute the $PAE\delta S(u,v)=|\{(t,uv) \in E \ : \ t \in [\Omega-\delta,\Omega] \}|$.
	Second, we take into account the activity of each pair of nodes between $\Omega$ and the occurrence time of the $k^{th}$ link between them before $\Omega$. 
	The corresponding index is $PAEkL(u,v)=k/(\Omega-t_{k})$ with $t_{k}$ such that $|\{(t,uv) \in L, \Omega\ge t\ge t_k\}| = k$.

	\subsubsection{Hybrid metrics}

	We also use hybrid metrics, which capture a mixture of structural and aggregated temporal information. 
	We use weighted variation of link prediction metrics as proposed in~\cite{tabourier2014rankmerging}.

	\begin{itemize}
	\item the \textit{Weighted Common Neighbors (\textit{WCN})} emphasize on the common neighbors which have often interacted with both nodes:
	
	\[WCN(u,v)=\sum_{w\in \mathcal{N}(u)\cap \mathcal{N}(v)}\mathcal{A}_{u,w} \cdot \mathcal{A}_{v,w} \]
	
	\item the \textit{Weighted Sorensen index (\textit{WSI})} is similar to the Sorensen index but take into account the activity of each node:
	
	\[WSI(u,v)=\frac {\sum_{w\in \mathcal{N}(u)\cap \mathcal{N}(v)}\mathcal{A}_{u,w} + \mathcal{A}_{v,w}}{\sum_{k \in \mathcal{N}(u)} \mathcal{A}_{u,k} + \sum_{k \in \mathcal{N}(v)} \mathcal{A}_{v,k}} \]
	
	\item the \textit{Weighted Adamic-Adar  (\textit{WAA})} decreases the weight of shared neighbors with high degree and high level of activity:

	\[WAA(u,v)= \sum_{w \in \mathcal{N}(u) \cap \mathcal{N}(v)} \frac{1}{\sum_{k \in \mathcal{N}(w)} \log(\mathcal{A}_{w,k})}\]		
	
	\item \textit{Weighted Resource Allocation (\textit{WRA})} similar to the Weighted Adamic-Adar but with giving weight differently:
	
	\[WRA(u,v)= \sum_{w \in \mathcal{N}(u) \cap \mathcal{N}(v)} \frac{1}{\sum_{k \in \mathcal{N}(w)} \mathcal{A}_{w,k}}\]

\end{itemize}

The metrics presented and their associated acronyms here are recalled in Table~\ref{Metricacronyms}.

\subsection{ Normalized metrics }

{
	Each metric is then normalized with the maximum value of the metric index. This allows for comparable metric index values and better understanding of each metric weight in the prediction.
	This normalization process is done for each pair of nodes and each metric in order to keep the absolute values comparable between different metrics.
	We obtain the normalized index $\hat{m}(u,v)$:
	
	\[\hat{m}(u,v) = \frac{ m(u,v)}{max_{(u,v)}  m(u,v)  }  \]
	
	Where $m(u,v)$ is the index associated to the metric $m$ over the pair of nodes $(uv)$.
}

\subsection{Prediction index}

{
	To use the information captured by the metrics presented above, we combine linearly the normalized metric indexes. 
	We build a prediction index $\mathcal{F}$, such that for all $uv\in V \otimes V$  $\mathcal{F}(u,v)$ represents how likely is the apparition of a link between $u$ and $v$ during $[A',\Omega']$.  
	Formally,
	
	\[ \mathcal{F}(u,v) =  \sum_{m\in \mathcal{M}} \alpha_m \cdot \hat{m}(u,v)   \]
	where $\mathcal{M}$ is the set of metrics used and $\hat{m}(u,v)$ is the normalized metric index associated to the metric $m$.
	The parameters $\alpha_m$ control each metric weight in the prediction index.
	%
	%
	Note that the important information does not hold in the absolute value of $\mathcal{F}(u,v)$, but rather in the relative values of $\mathcal{F}(u,v)$ in comparison to other $\mathcal{F}(u',v')$ as we will see in Section~\ref{protocole}.
	Note also that other combination methods are possible, the choice of linear combining being made for simplicity.
}

{
	Given such a prediction index, a standard method consists in learning on the training period values of $\alpha_m$ which optimize a given evaluation criterion. 
	Then, these weights are used for actual prediction on the prediction period. Our method to optimize the weights is presented in Section~\ref{learning}.
}

\subsection{Global prediction and link allocation}
\label{protocole}

{
	To predict the number of interactions between each pair of nodes during $T'=[A',\Omega']$, we first estimate the global number of links $N$ between all node pairs during this period as illustrated in Figure~\ref{extrapolation principle}. 
	We make the simplistic assumption that the global activity in $L'$ is the same as in $L$, and therefore, extrapolate linearly the stream activity to determine the global number $N$ of links to predict:
	$$	
		N=|E|\cdot \frac{\Omega'-A'}{\Omega-A}
	$$
	
	\begin{figure}[!h]
		\begin{center}
			\includegraphics[width=10cm]{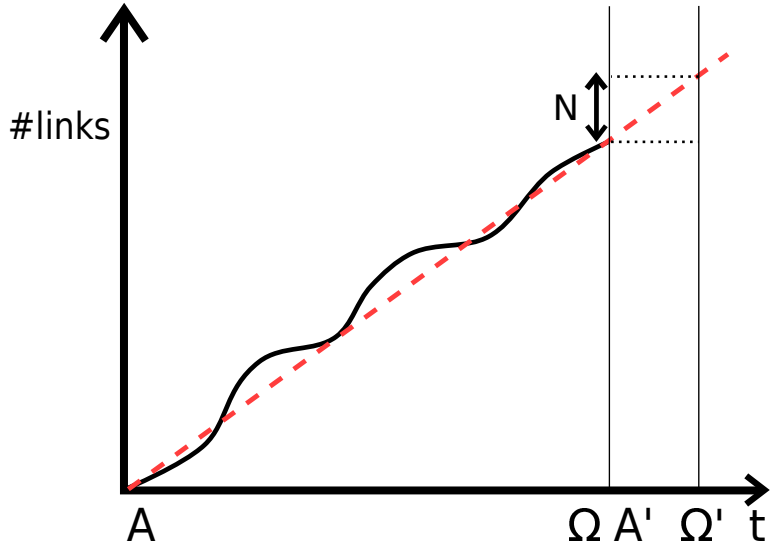}
			\caption{Global extrapolation principle representing the number of link having occurred during the input period. We obtain the number $N$ of link to predict during $[A',\Omega']$ by a linear extrapolation (dashed line).}
			\label{extrapolation principle}
		\end{center}	
	\end{figure}

}

{
	Then, for each pair of nodes, the metric indexes $m$ and subsequent normalized metric indexes $\hat{m}$ are computed on the link stream $L$. 
	As the prediction index reflects how likely the occurrence of a link between two nodes is, we use it to distribute the $N$ links between all the pairs in $V \times V$.
}	

{
	We allocate the $N$ links estimated previously proportionally to the prediction index to get $N_{(u,v)}$, the number of interactions predicted for any pair $uv$:
	
	\begin{equation}
	N(u,v)= N \cdot \frac{\mathcal{F}(u,v)}{\sum_{x,y \in V} \mathcal{F}(x,y)} 
	\end{equation}
	As mentioned previously, it is thus the relative values of the $\mathcal{F}(u,v)$ which are important in the computation of $N_{(u,v)}$, with $\sum_{x,y \in V}N_{(x,y)}=N$.
} 

{
	This framework allows to predict the future activity of the node pairs, that is the number of links appearing between each pair of nodes during $T'$.
	It is important to note that due to the specificities of our prediction task and in contrast to what is usually done for link prediction in graphs, this number is not necessarily an integer.}

\subsection{Evaluation protocol}

{
	To evaluate the efficiency of our protocol, we have to define a method suited to the prediction of multiple links for a pair of node which is also adapted to the fact that the number of predicted link is not a integer.

	Nevertheless, the evaluation method defined here aims to stay as close as possible to the tools used in classification tasks, which would make easier to adapt other prediction algorithms to the link activity prediction task in the future.
	We compare the set $\{N(u,v), uv \in V \otimes V\}$, with $N(u,v)$ the number of links predicted for the pair $uv$, with $\{N'(u,v), uv \in V \otimes V\}$, $N'(u,v)$ being the number of links that have actually occurred between $A'$ and $\Omega'$. 
	
	Thus, we adjust the usual definition of true positives, false positives and false negatives to the context of activity prediction in link streams, as these measurements underlie many of the evaluation metrics in prediction tasks.
	Precisely, for each pair $uv$, we compare $N(u,v)$, to $N'(u,v)$,(see Figure~\ref{eval}). 
	We then define the number of \textit{TP}, \textit{FP} and \textit{FN} as follows:
	
	$$\left\{
	\begin{array}{l}
	TP(u,v) = \min(N(u,v),N'(u,v))\\
	FP(u,v) =  \max(N(u,v)-N'(u,v),0)\\
	FN(u,v) =  \max(N'(u,v)-N(u,v),0)
	\end{array}
	\right.$$

	The sum of each of these indicators over all node pairs yields the number of $TP$, $FP$ and $FN$ for all the predictions. 
	Note that these definitions allow to get the usual relationships between the indicators, in particular, $TP + FP$ is the number of predictions and $TP + FN$ is the total number of interactions actually occurring during $T'$.
	Moreover, they convey the same idea as the usual \textit{TP}, \textit{FP} and \textit{FN} do, as \textit{TP} and \textit{FP} reflect respectively the number of good and false predictions among all, while \textit{FN} corresponds to actual occurrences of interactions which have not been predicted.

	Consequently, we can compute more sophisticated performance indicators, like the precision $ \frac{TP}{TP+FP} $,and the recall $ \frac{TP}{TP+FN} $. 
	We also use the F-score to quantify the quality of prediction, which is the harmonic mean of these two indicators: $2 \cdot \frac{ precision \cdot recall}{precision+recall}$. 
	Other indicators could be defined in this context, like the ROC curve, but we do not use them in this study.
	
	\begin{figure}[h]
		\begin{center}
			\includegraphics[width=10cm]{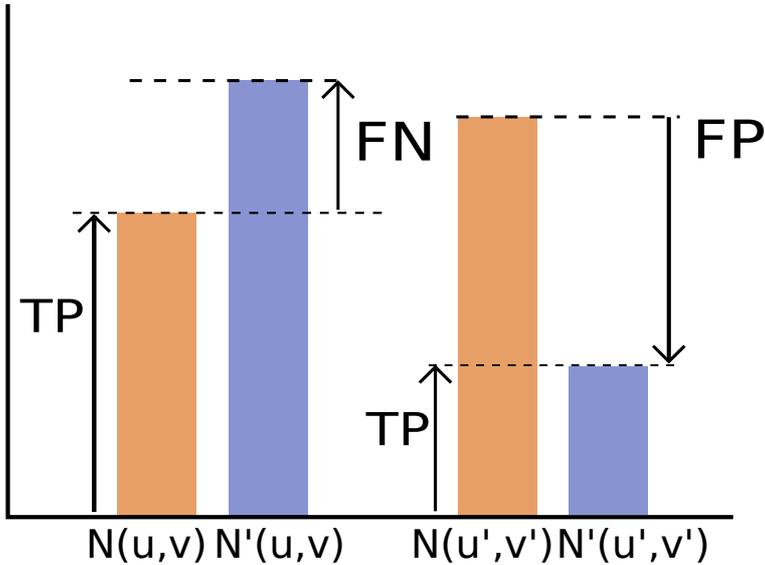}
			\caption{Examples of evaluation scores for activity prediction in link streams with too low prediction (left) and too high prediction (right)}
			\label{eval}
		\end{center}
	\end{figure} 
}

\section{Learning}
\label{learning}

{
	The choice of values for parameters $\alpha_m$ is a crucial step to achieve accurate prediction, as they determine the balance between combined metrics. 
	An automated way to choose these values is also needed when using multiple metrics simultaneously, as systematically explore the parameter space would be too costly.
	For this purpose we use a machine learning algorithm and define training and testing phases. 
}	

\subsection{ Training, validation, observation and prediction streams}

We use a hold-out technique, which means that we divide the data into sub-streams in order to define a training phase during which we optimize the parameters values and then extrapolate these parameters for the purpose of the prediction itself.
Note that $k$-fold validation is much used in prediction tasks, however it is difficult to implement in a context where the time-ordering plays an essential role, as using randomly chosen periods of time would lead to an important loss of temporal structure.

\begin{figure}[h]
	\begin{center}
	\includegraphics[width=10cm]{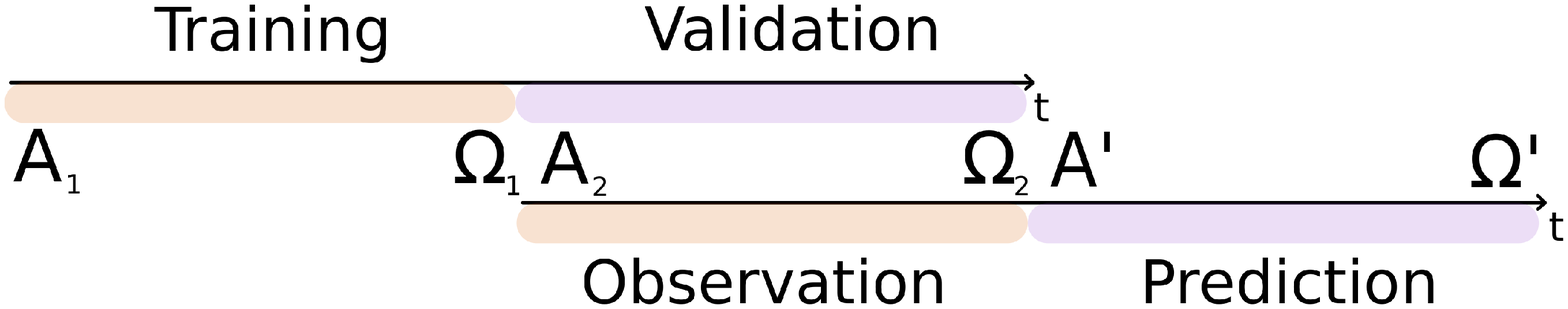}
	
	\caption{Illustration of the different periods used in our protocol: the training and validation periods (top) used to train the algorithm and the observation and prediction periods (bottom) used to make the actual prediction.}
	\label{shemaperiodes}
\end{center}	
\end{figure}

{
	Practically, we proceed in the following way, see Figure~\ref{shemaperiodes}:
	we divide the input stream $L$ into two sub-streams: a \textit{training sub-stream} $L_1=(T_1,V,E_1)$ with $T_1=[A_1,\Omega_1]$, used to compute the metrics during the training phase, and a \textit{validation sub-stream} $L_2=(T_2,V,E_2)$ with $T_2=[A_2,\Omega_2]$.
	The values of parameters $\alpha_m$ are then computed using a learning algorithm to optimize activity prediction on $T_2$ using the information contained in $T_1$.
	Prediction metrics are then computed on the actual observation sub-stream $L_2=(T_2,V,E_2)$ with $T_2=[A_2,\Omega_2]$, to predict the activity during the prediction stream $L'=(T',V,E')$ with $T'=[A',\Omega']$ .
}

In our implementation, the validation and the effective observation sub-streams are identical.
This choice is not mandatory, but we make it because it would probably be the operator choice in many real-time applications, as it means that the latest observations are used in order to predict coming events.

\subsection{Gradient descent}

{
	We use a gradient descent algorithm to explore the parameter space and find good values for each parameter $\alpha_m$. 
	Using a linear combination of metric indexes to avoids repeating a number of heavy computational steps. 
	The metrics do not need to be recomputed for each combination.
	Indeed, as we combine metric indexes for each pair of nodes, this approach allows to predict activity using different sets of parameters without any additional computation. 
	This reduces greatly the computing cost of the method.
	The initial set of parameters given as input to the learning algorithm is drawn randomly in the parameter space for each prediction.
}

\section{Method Summary}

\begin{figure}[h!]
	\begin{center}
	\includegraphics[width=9cm]{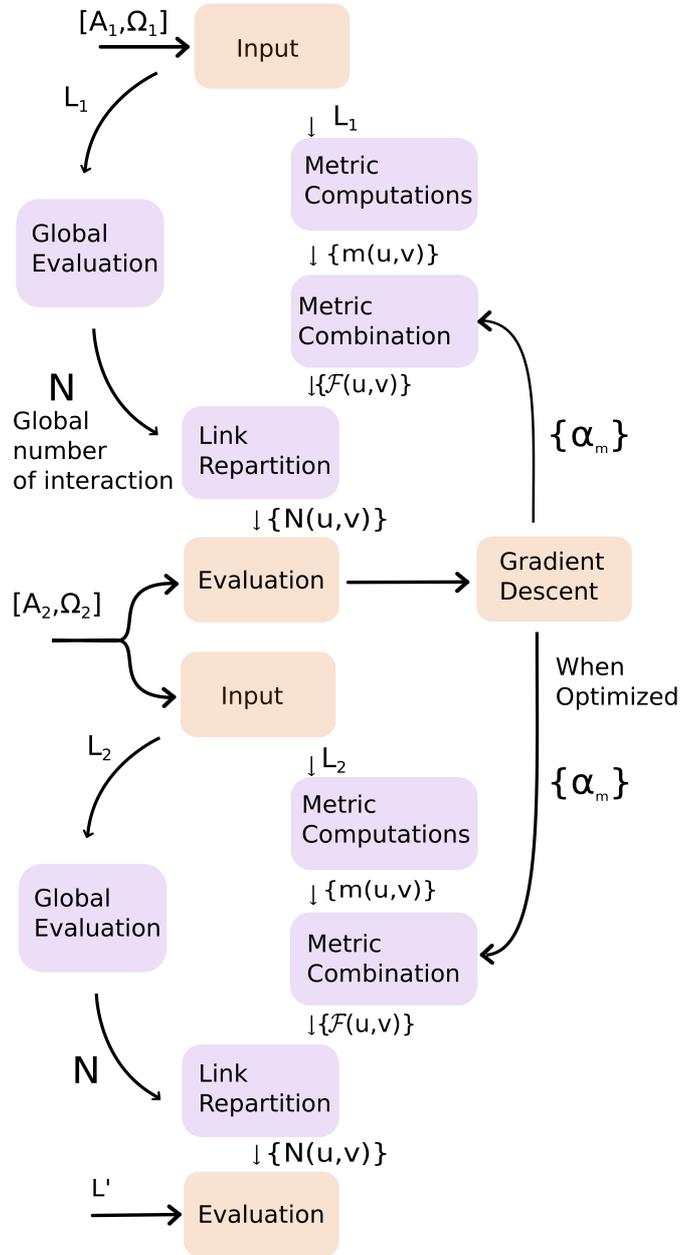}
	
	\caption{Summary of our prediction protocol, with learning and evaluation phases.}
	\label{shemaalgo}
	\end{center}											 
\end{figure}

Our implementation is available at https://github.com/ThibaudA/linkstreamprediction.	
Our protocol is able to capture different information and to combine them to optimize prediction. 
We insist on the fact that it is built as much as possible in a modular way, in order to allow others to easily adapt the protocol to specificities of a given activity prediction problem.
We represent the different steps of our approach in Figure~\ref{shemaalgo}.

\section{Datasets description}
\label{datadescription}

In order to assess the performances of our framework, we conducted experiments on four datasets which gather contacts detected with sensors. 
Each undirected link $(t, uv)$ means that the sensor carried by nodes $u$ or $v$ detected the sensor carried by the other nodes at time $t$, which means in turn that these two nodes were close enough at time $t$ for the detection to happen. 
We call this a contact between nodes $u$ and $v$.
We present the datasets briefly in this section.

{

	 The first trace was collected in a French high school in 2012 (\textit{Highschool} dataset), see \cite{mastrandrea2015contact} for full details. 
		It is a link stream of 181 nodes and 45047 links, connecting 2220 distinct pairs of nodes over a period of approximately 8 days. 

		The second dataset has been collected during the IEEE INFOCOM 2006 in Barcelona (\textit{Infocom} dataset) -- see~\cite{cambridge-haggle-20090529}. 
		The bluetooth devices used in this experiment recorded connexions with one another. 
		This dataset contains 98 nodes and 283,100 links. During this 3 days long experiment, 4338 pairs of nodes have interacted.
		Note that the \textit{Infocom} dataset, which is also a contact sensor trace, involves less nodes but contains more links and more active node pairs than the \textit{Highschool} dataset does.

		Our third dataset is the \textit{Reality Mining} dataset, which gather contacts among students of the Massachusetts Institute of Technology as recorded by their mobile phones -- see~\cite{mit-reality-20050701}.
		The trace contains 1,063,063 links between 96 nodes during 9 months with 2539 distinct pairs of nodes involved in interactions.

		Finally, we study the \textit{Taxi} dataset. 
		It is based on the GPS locations of 305 taxis in Rome, recorded during the month of February 2014 -- see~\cite{roma-taxi-20140717}.
		We consider that two taxis have interacted if they are within a 30 meters range from each other. 
		This yields a link stream of 22,364,061 links. During the experiment 16,799 pairs of nodes have interacted, which makes this dataset significantly larger than the other datasets under study.

}

\section{Experiments}

In this section, we study the results of our activity prediction method on the datasets previously described. 
First, we implement the standard prediction protocol aforementioned.	
Then, we investigate more closely how the learning method distributes weights among combined metrics.
Finally, we identify the strengths and weaknesses of the procedure and important challenges to improve the activity prediction.

\subsection{Experimental implementation}

{	
	
	As a reminder of the protocol described in Section~\ref{learning}, we divide our input period $L$ in two substreams, the training stream $L_1=(T_1,V,E_1)$ with $T_1=[A_1,\Omega_1]$, and the validation and observation stream $L_2=(T_2,V,E_2)$ with $T_2=[A_2,\Omega_2]$.
	Values of parameters $\alpha_m$ are then chosen using a gradient descent algorithm that optimize activity prediction on $T_2$ using the information contained in $L_1$.
	The prediction metrics are then computed on the observation sub-stream $L_2=(T_2,V,E_2)$ with $T_2=[A_2,\Omega_2]$, to predict the activity in prediction stream $L'=(T',V,E')$ with $T'=[A',\Omega']$.
	
	In our experimental set-up the durations of training, validation, observation and prediction periods are chosen equal. 
	In the cases of \textit{Highschool} and \textit{Infocom} datasets, we use 1, 2 and 3 hours long periods starting on a Monday at 8:30 and 9:00 respectively. 
	In the cases of \textit{Reality Mining} and \textit{Taxi} datasets, the timescales are typically longer, so we use 1, 2 and 3 days long periods, starting on a Tuesday at 1:30 am for the \textit{Reality Mining} dataset and a Wednesday at 8:00 am for the \textit{Taxi} dataset.

	We report in Table~\ref{tabletime} the exact starting and ending times for \textit{Highschool} and \textit{Infocom} experiments. The acronyms used in the following are indicated in Table~\ref{Metricacronyms}.
	The choice of these periods have important consequences on the prediction quality, as we shall see in the next section. 	

}

\begin{table}[h!]
	\begin{center}
		\caption{Metric Acronyms}

		\begin{tabular}{ l  l  }
			
			\hline
			Acronym & Metric   \\ \hline 
			CN & \textit{Number of Common Neighbors} \\ \hline 
			AA & \textit{Adamic-Adar Index}   \\ \hline 
			RA &  \textit{Resource Allocation Index}   \\ \hline 
			SI &  \textit{Sorensen Index}    \\ \hline 
			JI &  \textit{Jaccard Index}   \\ \hline 
			WCN & \textit{Weighted Number of Common Neighbors}   \\ \hline 
			WAA & \textit{Weighted Adamic-Adar} \\ \hline 
			WRA &  \textit{Weighted Resource Allocation Index} \\ \hline 
			WSI &  \textit{Weighted Sorensen Index} \\ \hline 		
			PAE &   \textit{Pair Activity Extrapolation} \\ \hline 
			PAE10L  & \textit{Activity by unit of time during the last 10 links}  \\ \hline
			PAE100S      &  \textit{Activity during the last 100 seconds} \\ \hline
			PAE1000S	 &  \textit{Activity during the last 1000 seconds} \\ \hline
			PAE10000S	 &  \textit{Activity during the last 10000 seconds} \\ \hline
		\end{tabular}
	
		\label{Metricacronyms}
	\end{center}	
\end{table}		

\begin{table}[h!]
	\begin{center}
		\caption{\label{tabletime}Starting and ending time of each periods for each dataset}
		\begin{tabular}{| l |c |c | c | c | c|}
			
			\hline
			Dataset& Duration & $A_1$ & $\Omega_1 = A_2$  & $\Omega_2 = A'$& $ \Omega' $ \\ \hline 
			\multirow{3}{*}{Highschool}	  & 1h  & 8:30 am & 9:30 am    & 10:30 am   & 11:30 am      \\ 
			& 2h  & 8:30 am & 10:30 am    & 12:30 am  & 2:30 pm     \\ 
			& 3h  & 8:30 am & 11:30 am    & 2:30 pm  &  5:30 pm   \\ \hline 
			\multirow{3}{*}{Infocom} & 1h  & 9:00 am & 10:00 am    & 11:00 am &  12:00 am   \\ 
			& 2h  & 9:00 am & 11:00 am   & 1:00 pm   & 3:00 pm    \\ 
			& 3h  & 9:00 am & 12:00 am   & 3:00 pm & 6:00 pm     \\ \hline
			\multirow{3}{*}{Reality Mining} & 1d  & Tuesday& Wednesday    & Thursday &  Friday   \\ 
			& 2d  & Tuesday & Thursday  & Saturday   & Monday    \\ 
			& 3d  & Tuesday & Friday   & Monday & Thursday \\ \hline 
			
			\multirow{3}{*}{Taxi} & 1d  & Wednesday& Thursday    & Friday &  Saturday   \\ 
			& 2d  & Wednesday & Friday  & Sunday   & Tuesday\\ 
			& 3d  & Wednesday & Saturday   & Tuesday & Friday \\ \hline 
		\end{tabular}
	\end{center}
\end{table}

\subsection{Standard prediction}
\label{exp}

{
	Results are summarized in Table~\ref{("Res C0")}. 
	We report the F-score, the recall and the precision for each experiment, as well as the number of predicted links and the number of links which actually occurred during the prediction period.
	The values presented are the averages obtained on 10 realizations for each set of parameters. The standard deviation for F-score, precision and recall is in all cases lower than $ 0.01 $.
}

\begin{table}[h!]
	\begin{center}	
		\caption{\label{("Res C0")}F-score and number of link predicted and appeared}
		\begin{tabular}{| l |c |c|  c|c| c | c | }
			
			\hline
			Dataset& Duration & F-score &Prec. &Recall & Pred  & App  \\ \hline 
			\multirow{3}{*}{Highschool} & 1h  & 0.44 &  0.33 & 0.69 & 1123    & 857  \\  
			& 2h  & 0.17 & 0.18  & 0.15  & 1751    & 2178    \\  
			& 3h  & 0.29 & 0.22 & 0.41 & 3072    & 1900    \\ \hline 
			\multirow{3}{*}{InfoCom}& 1h  & 0.59 & 0.58 & 0.59  & 8167    & 8220    \\  
			& 2h  & 0.55 & 0.50 & 0.60 & 16737   & 14051   \\  
			& 3h  & 0.67 &  0.63 & 0.70 & 22568   & 20850   \\ \hline 
			\multirow{3}{*}{R.Mining}& 1d  & 0.47 & 0.41  & 0.56 & 6105    & 8733    \\  
			& 2d  & 0.09 & 0.05  & 0.65 & 18537   & 1591    \\ 
			& 3d  & 0.41 &  0.73 & 0.29  & 11395   & 29078   \\ \hline 
			\multirow{3}{*}{Taxi}   & 1d  & 0.18 &  0.19  & 0.17 & 872173  & 971204  \\  
			& 2d  & 0.10 & 0.07 & 0.16 & 1904076 & 910457  \\  
			& 3d  & 0.16 & 0.20  & 0.14  & 1843329 & 2865449 \\ \hline 
		\end{tabular}
	\end{center}
\end{table}

\subsubsection{Highschool}	

{	
	
	We first focus on the \textit{Highschool} dataset experiments. 
	We can see that during the 1 hour experiment, the algorithm predicts slightly more links than what actually appear. 
	Still, recall is $0.69$, meaning that 69\% of interactions observed have been correctly predicted, therefore the method seems to quite precisely accounted for a large part of the activity in the dataset during this period.

	However, a significant drop in the prediction quality can be observed during the 2 hours experiment, which is visible on both the precision and the recall (and consequently on the F-score).
	This is probably due to the fact that the observation period spreads over on lunch break.
	The overall activity predicted over the prediction link stream underestimates the number of links that actually appear.
	It reflects that the activity from the observation stream is not as high as the actual activity from the prediction stream.
	But the main reason for the drop in prediction quality certainly stems from the fact that the lunch break is an opportunity for interactions which differ from the ones occurring during class hours, and this seems to have a significant impact.
	We explore this question in more details in Section~\ref{badpred}.

	Considering the 3 hours experiment, we observe a slight increase in precision and recall compared to the 2 hours experiment. 
	We think that longer observation periods mitigate the effect of behavior changes, like the one that happens at lunch break.
	The overall activity predicted is greater that the activity that actually occurs. 
	While this effect impacts the prediction quality, we see that it is not as significant as with the 2 hours experiment.

	These experiments show that predictions on longer periods are negatively affected by the strong variation in the behavior of the system over time, mostly due to the high-school schedule, which alternates breaks and class hours.
}

\subsubsection{Infocom}	

{
	For the \textit{Infocom} dataset, performances are more stable compared to the \textit{Highschool} case,  in particular when comparing the number of predicted links to the number of occurring links. 
	Indeed, the divergence between these values remains consistently lower than 20\%.
	This leads to better performances on each experimentation. 
	We suggest that, as this dataset reports contacts at a conference, it may be explained by the fact that differences in behaviors are less marked between talks and breaks. 
	At a much smaller scale than in the \textit{Highschool} case, we also observe a loss in prediction quality when using two hours periods compared to 1 hour periods, but the effect disappears on longer periods: 3 hours training and prediction periods yield even better results than the 1 hour case. 
}

\subsubsection{Reality mining}	

{
	The \textit{Reality Mining} dataset performs relatively well on 1 day and 3 days periods, however we can see a dramatic loss in prediction quality using 2 days periods.} 
In this case, the loss seems to come from the vast overestimate of predicted links compared to the links which appear.
This poor performance prediction is investigated more thoroughly in Section~\ref{badpred}.

In the case of the 3 days training and prediction periods experiment, the overall predicted activity is now lower than the occurring activity, which leads to a low recall value ($0.29$) relatively to the 1 day experiment ($0.56$), where the underestimation of activity was not as significant.
However, predicted interactions are quite accurate, as the precision is $ 0.73 $.
This tends to shows that in certain instances our protocol is able to achieve suitable prediction despite an underestimated activity during prediction period.

\subsubsection{Taxi}

Considering the \textit{Taxi} dataset, we can see that prediction quality in terms of F-score is lower than for other datasets.
{
	This is due to a greater number of nodes and interactions in the dataset, leading to a more difficult prediction task, as there are more candidate node pairs which can be predicted.
	Notice also that the nature of the dataset is different and geographical proximity between two taxis may be less significant that proximity of students or conference attendees.} 

The 1 day experiment predicts an overall activity closer to the actual activity, leading to the best prediction. 
We can see that for 2 days and 3 days predictions, the extrapolation yields less accurate results regarding the F-score, and in each case the overall activity prediction has been largely ill-estimated (respectively over and underestimated).

{
	These experiments show that the choice of the prediction period plays a key role for prediction quality. 
	It is closely related to the overall activity extrapolation, leading to important variations in the prediction quality.  
}

\subsection{Metric combinations}
\label{sec:feat-comb}

\begin{figure}[!h]
	\begin{center}
		
		\includegraphics[width=9cm]{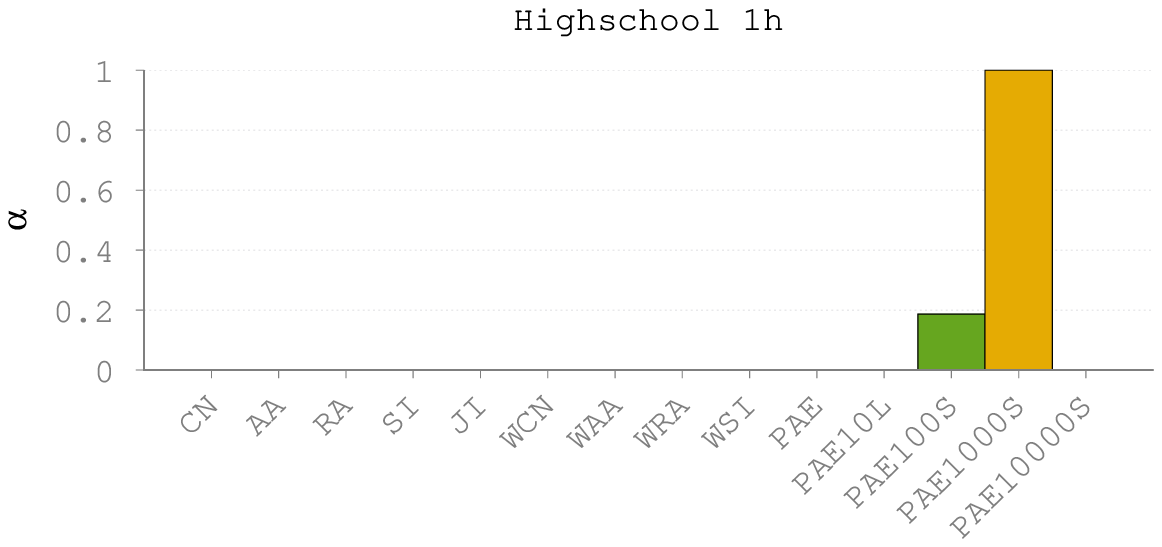}
		\includegraphics[width=9cm]{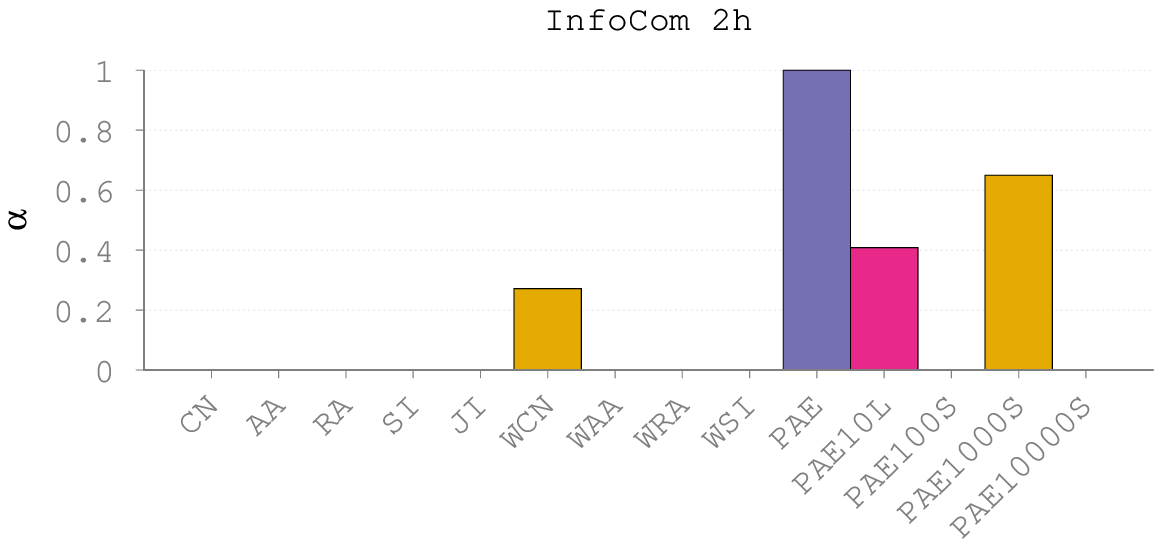}
		\includegraphics[width=9cm]{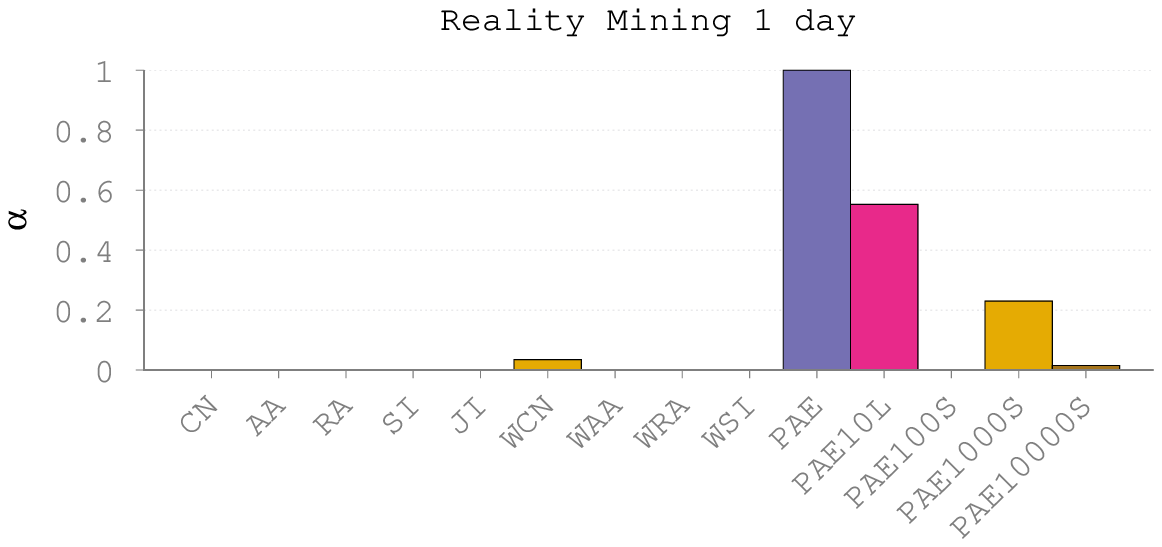}
		\includegraphics[width=9cm]{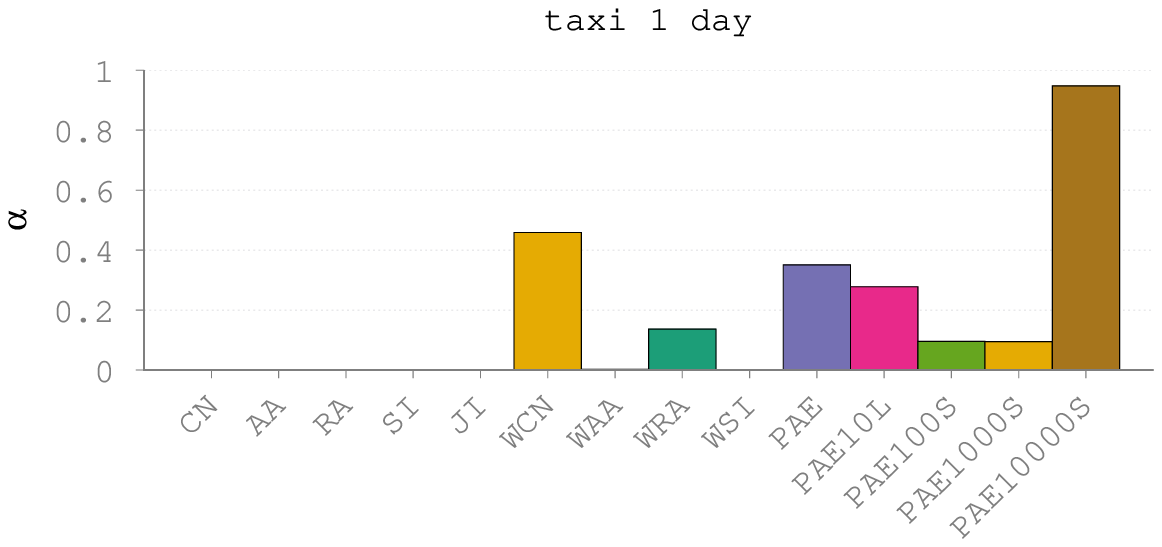}
		\label{("Mix C0")}
		\caption{Metric coefficients repartition computed by the learning algorithm for the following datasets and training periods (from top to bottom): \textit{Highschool} (1h), \textit{InfoCom} (2h), \textit{Reality Mining} (1d) and \textit{Taxi} (1d).}
		
	\end{center}
\end{figure}

{
	In this section we observe how the algorithm mixes different metrics in order to optimize prediction. 
	The combination realized for each dataset is shown in Figure~\ref{("Mix C0")}. 
	The height of each bar corresponds to the mean of coefficients obtained for the experiments presented above (average on 10 realizations for each set of parameters). }

{
	We have represented metric combinations for the following training/prediction periods and datasets: 1 hour for \textit{Highschool}; 2 hours for \textit{Infocom} and 1 day for \textit{Reality Mining} and \textit{Taxi}. 
	The meaning of metric acronyms used can be found in Table~\ref{Metricacronyms}.
}

{
	Considering the \textit{Highschool} dataset, algorithm main\-ly uses the \textit{Pair Activity Extrapolation} (PAE) metric with a small influence from the \textit{activity during the last 100 seconds} (PAE100S). 
	This tells us that the algorithm basically extrapolates the dynamics of each pair of nodes during the end of the observation period to predict future interactions.}
Considering the \textit{Infocom} dataset, the algorithm also focuses on temporal metrics.
{
	It allows the algorithm to gather information about dynamics related at different time scales. 
	The algorithm also makes use of one of the weighted metric, namely the \textit{Weighted number of Common Neighbors} (WCN).}
%

%
The patterns of used metrics in the case of the \textit{Reality Mining} is quite similar to the \textit{Infocom} case.
{
	Indeed, the algorithm gives most weight to temporal metrics, with a marginal use of \textit{Weighted number of Common Neighbors}, more precisely the \textit{activity during the last 1,000 seconds} (PAE1000S) and the \textit{activity by unit of time during the 10 last links} (PAE10L). 
	Again the time granularity being 120 seconds, \textit{activity during the last 100 seconds} does not appears.
}

{
	Finally, concerning the \textit{Taxi} dataset, we can see that our algorithm mainly uses \textit{activity during the last 10,000 seconds} (PAE10000S).} 
However, all the other temporal metrics are also used with relatively lower weights, indicating that all timescales carry a certain amount of information in regards to this prediction.
Besides that, we also see a notable contribution of structure-oriented index such as the \textit{Weighted number of Common Neighbors} as well as a slight use of the \textit{Weighted Resource Allocation index}.

{
	For each dataset the algorithm uses slightly different metric combinations. 
	However, we can note that temporal metrics are often favored. 
	Purely structural metrics are mostly ignored by our algorithm, yet hybrid metrics collect structural information, which is used by the algorithm to improve the prediction in most experimental settings.
	It is also interesting to note that temporal metrics corresponding to different time scales are represented, meaning that combining these metrics allows the algorithm to capture different information.
}

	An important point is that by extrapolating the past of a pair of nodes, temporal metrics are simply not able to predict new interactions, while structural and hybrid one may.
This indicates that it is much harder to predict interactions that appear for the first time than previously observed interactions.
Indeed, our framework combines metrics in order to focus on repeated links rather than on new links.
We explore this intuition thereafter.

\subsection{The case of poor predictions}
\label{badpred}

	In this section, we investigate in more details the two cases seen in Section~\ref{exp} which exhibit the lowest prediction quality in terms of F-score, compared to other experiments on the same dataset.

\begin{figure}[h!]
		\begin{center}
			
			\includegraphics[width=12cm]{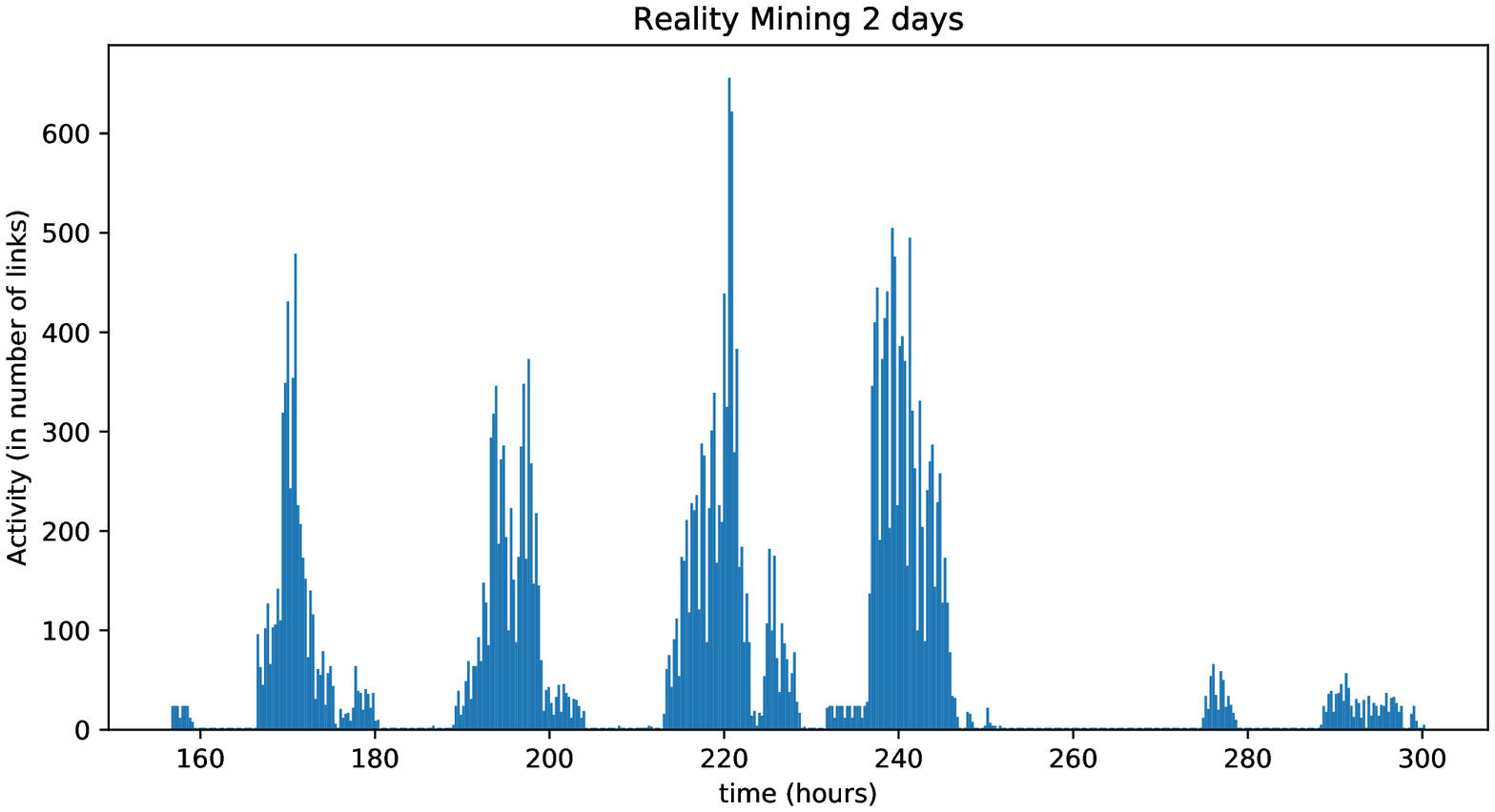}
			\includegraphics[width=12cm]{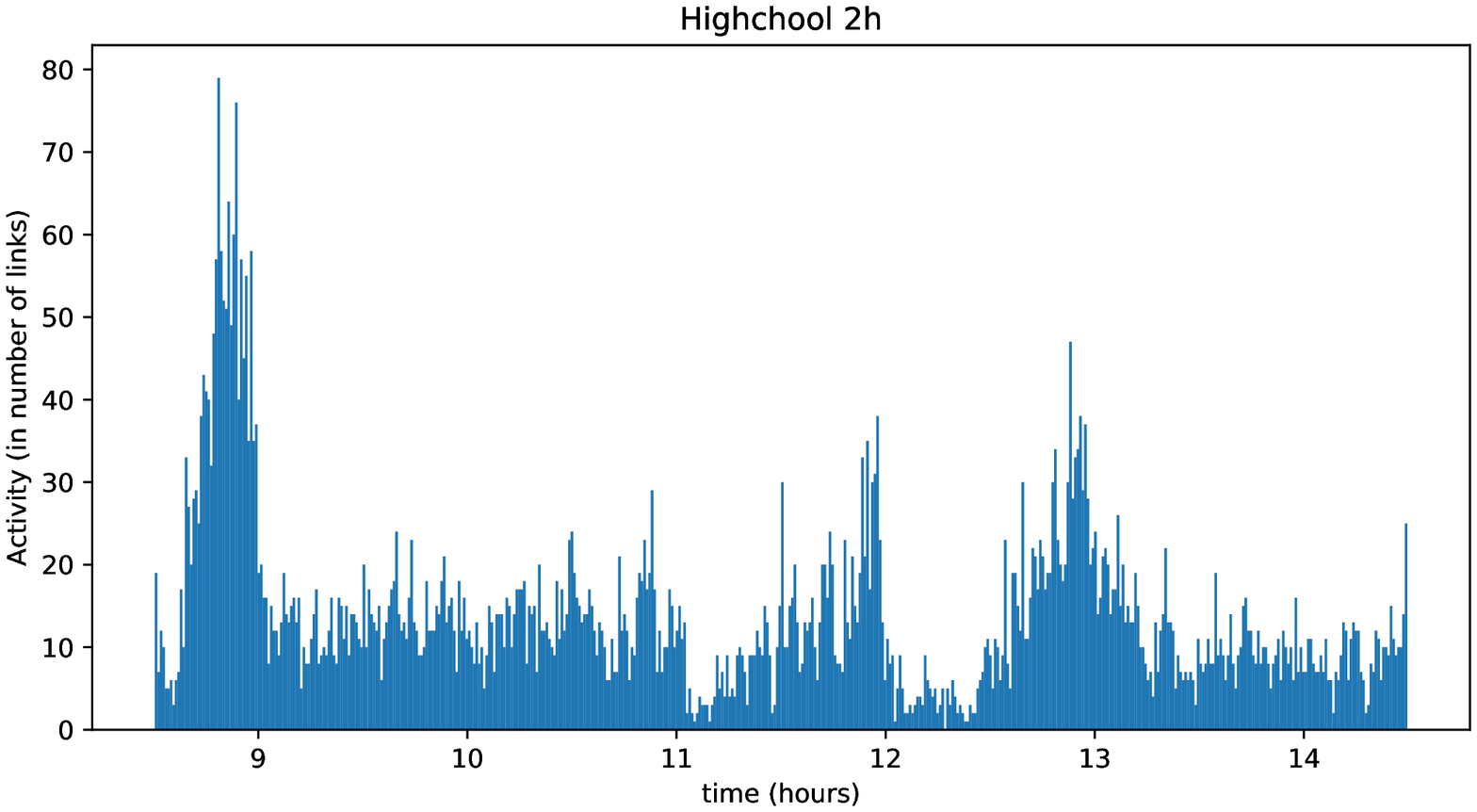}

			\caption{\label{linkapp} Number of links appearing over time during \textit{Reality Mining} 2 days experiment (Granularity: 1000 seconds) and \textit{Highschool} 2 hours experiment (Granularity: 40 seconds).}
			
		\end{center}
	\end{figure}

	Considering the 2 days experiment on \textit{Reality Mining}, we have observed an important drop of the overall activity during the prediction period. 
	We show in Figure~\ref{linkapp} the number of links through time during this experiment with a granularity of $1000$ seconds. 
	The system displays a large decrease on the number of links appearing during the last two days. 
	The experiment starting on a Tuesday, this mean that the  training period is from Tuesday to Wednesday, the validation and observation from Thursday to Friday, while the prediction period is Saturday and Sunday, which leads to a huge decrease of activity.

		Concerning the \textit{Highschool} experiment with 2 hours periods, we have previously noted an underestimate of the activity during the prediction period. Again, we show in Figure~\ref{linkapp} the activity during this experiment with a granularity of $40$ seconds. 
		We have also seen in Table~\ref{tabletime} that the prediction period $[A',\Omega']$ spans from 12:30 pm to 2:30 pm. 
		This corresponds to lunch break, when students leave their classes and have the opportunity to mix with other students. 
		It is hard to draw any conclusion from Figure~\ref{linkapp}, as the activity decreases from the training period (8:30 am to 10:30 am) to the validation period (10:30 am to 12:30 pm) and again from the validation period to the prediction period.
		But the predicted overall activity reported in Table~\ref{("Res C0")} seems not far enough from the actual activity to explain alone the drop in prediction quality. 
		To investigate further this problem, we achieve a prediction task,  which is not realistic but allows us to have a better grasp of the reason for the former observation.
		We repeat the previous experiment, but setting predicted activity to actual activity measured in the prediction stream. 
		This experiment yields a F-score of $0.17$, precision is $0.16$ and recall is $0.17$. These values are close to the quality obtained during the original experiment.
		This shows that predicted overall activity alone does not explain the drop in quality. 
		Thus, the prediction is mostly affected by the change of students contact behavior from classes to lunch time.

	Notice that these results evidence the importance of the choice of the training, validation and prediction periods in the protocol.

	\section{Investigating combined metrics}

	In this section, we use a simplified learning model, identical to the one presented above except for the fact that it mixes only two different metrics.
	This simplification allows to clearly represent the mixing achieved by the method, depending on the values of learning parameters.
	Our purpose is to measure what kind of links are predicted depending on the mixing.
	To do so, we focus on the \textit{Highschool} and \textit{Infocom} datasets.
	As discussed later, this will give us additional hints about the relatively low contribution of structural metrics in previous series of experiments.

	\subsection{Simplified experimental setting}

		The experimental protocol focuses on the performance gain that can be achieved by combining two link stream metrics. 
		Here, we focus on the training and validation periods only. 
		Then, to understand the information brought by each metric, we combine two metrics at a time, the weight of each metric being related to the parameter $\alpha$, according to the following equation:
		\begin{equation}
		\label{combexp}
		\mathcal{F}(u,v)=\alpha \cdot \widehat{PAE1000S}(u,v)+ (1-\alpha) \cdot \hat{m}(u,v)
		\end{equation}
		where $PAE1000S$ is the index associated to the \textit{activity during the last 1000 seconds} (see~\ref{tempmetrics}).
		We use it as a reference because we have seen that it is the most weighted metric in the combination in the \textit{Highschool} dataset and the second most weighted metric in the \textit{Infocom} dataset. 
		For each experiment we combine the \textit{activity during the last 1000 seconds} with $\hat{m}(u,v)$ the index corresponding to an other metric, as indicated in the legend of each figure.

	\subsection{Structural and temporal metrics combinations}

\begin{figure}[!h]
		\begin{center}
			
			\includegraphics[width=12cm]{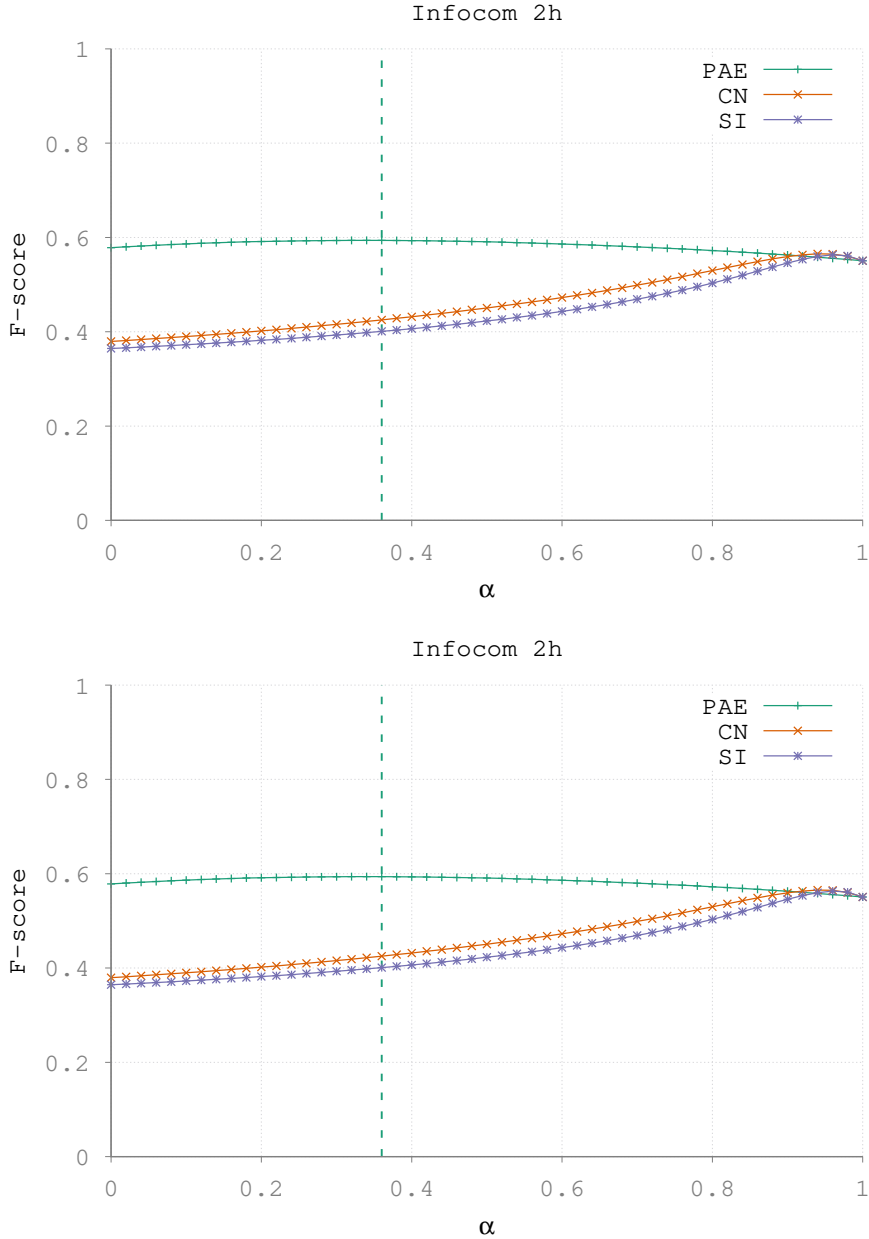}
			\includegraphics[width=12cm]{3metriques.eps}
			\caption{F-score of the predictions with different weight ratios between the \textit{activity during the last 1000 seconds} and three other metrics for \textit{Highschool} (1h) (top) and between the \textit{Pair activity extrapolation} and three other metrics for \textit{Infocom} (2h) (bottom). Dashed lines indicate the ratio between the two main metrics during the experiment in Section~\ref{sec:feat-comb} (in green in both figures).}
			\label{3metrics} 
		\end{center}
	\end{figure}

		We study how the use of different metrics with different weights affects prediction on \textit{Highschool} and \textit{Infocom} datasets.
		For this purpose, we combine three of the metrics presented in section~\ref{structmetrics} with the \textit{activity during the last 1000 seconds}. 
		For each dataset we combine it to the most important metric index used by our learning algorithm, that is to say the \textit{activity during the last 100 seconds} for the \textit{Highschool} dataset and the \textit{Pair Activity Extrapolation} for the \textit{Infocom} dataset.
		We also combine it with the \textit{number of Common Neighbors} and the \textit{Sorensen Index} to study how the algorithm mixes \textit{activity during the last 1000 seconds} to structural metrics. 
		Plots in Figure~\ref{3metrics} represent the F-score for different pairs of metric combinations, as a function of $\alpha$ for each of the two datasets. 
		Higher values of $\alpha$ represent a greater weight of the \textit{activity during the last 1000 seconds} in the prediction.

		In the case of \textit{Highschool}, we use a 1 hour training period.
		We combine successively \textit{PAE1000S} with \textit{PAE100S}, the \textit{number of Common Neighbors} (CN) and the \textit{Sorensen Index} (SI).
		The plots for \textit{PAE100S} show that F-score increases with $ \alpha $ until reaching a maximum for $\alpha \simeq 0.74$ and then decreases until $ \alpha =1 $. 
		With the structural metrics we observe roughly similar behaviors: a slow increase until $\alpha \simeq 0.8$ then a rapid increase until a maximum at $\alpha = 1$. 
		This shows that the algorithm simply does not benefit at all from the structural metrics in this experiment.
		Indeed, when $\alpha=1$, the prediction index comes down exactly to the \textit{activity during the last 1000 seconds}.

		We then apply the same protocol to \textit{Infocom}, with a 1 hour training period. 
		This time we combine \textit{PAE1000S} with the \textit{Pair Activity Extrapolation}, \textit{CN} and \textit{SI}, successively. 
		The mixing between temporal metrics (\textit{PAE1000S} and \textit{PAE}) behaves qualitatively in the same way as the mixing of temporal metrics in the \textit{Highschool} case, that is a convex curve with a maximum for $\alpha \simeq 0.35$.
		The combination with structural metrics behave qualitatively in a different way than in the \textit{Highschool} case. 
		Both plots have similar shapes, showing an increase until a maximum for $ \alpha \simeq 0.94 $ then decreasing until $\alpha=1$.
		These observations indicate that in this experiment, combining a temporal metric with a structural metric may lead to an improvement of the F-score.

		Observations of this section confirm that the algorithm is able to find a trade-off between metrics that improve the performance of the prediction.
		Moreover, we see more clearly that the mixing tends to heavily favor temporal metrics.
		We will now turn our attention to the kind of links which are consequently predicted.

		\subsection{Nature of predicted links}
		\label{Categories analysis}

			Using the same prediction protocol, we divide the set of node pairs into two categories, to see what kind of links are predicted in each case.
			On the one hand, some pairs have not interacted during $T$, so that predicting the occurrence of a link between nodes of this kind is predicting a new link in the stream.
			We call \textit{new link} any $(t,uv)$ in the prediction stream $L'$ such that $\nexists \ x \in T, (x,uv) \in E$.
			On the other hand, other pairs have interacted during $T$ and predicting such an interaction is predicting the repetition of a link. 
			We call \textit{recurrent link} any $(t,uv)$ in the prediction stream $L'$ such that there exists a link $(x,uv) \in E$.
			We apply the evaluation method on the complete set of pairs and on each of these two subsets. 

		\begin{figure}[!h]
			\begin{center}
				\includegraphics[width=12cm]{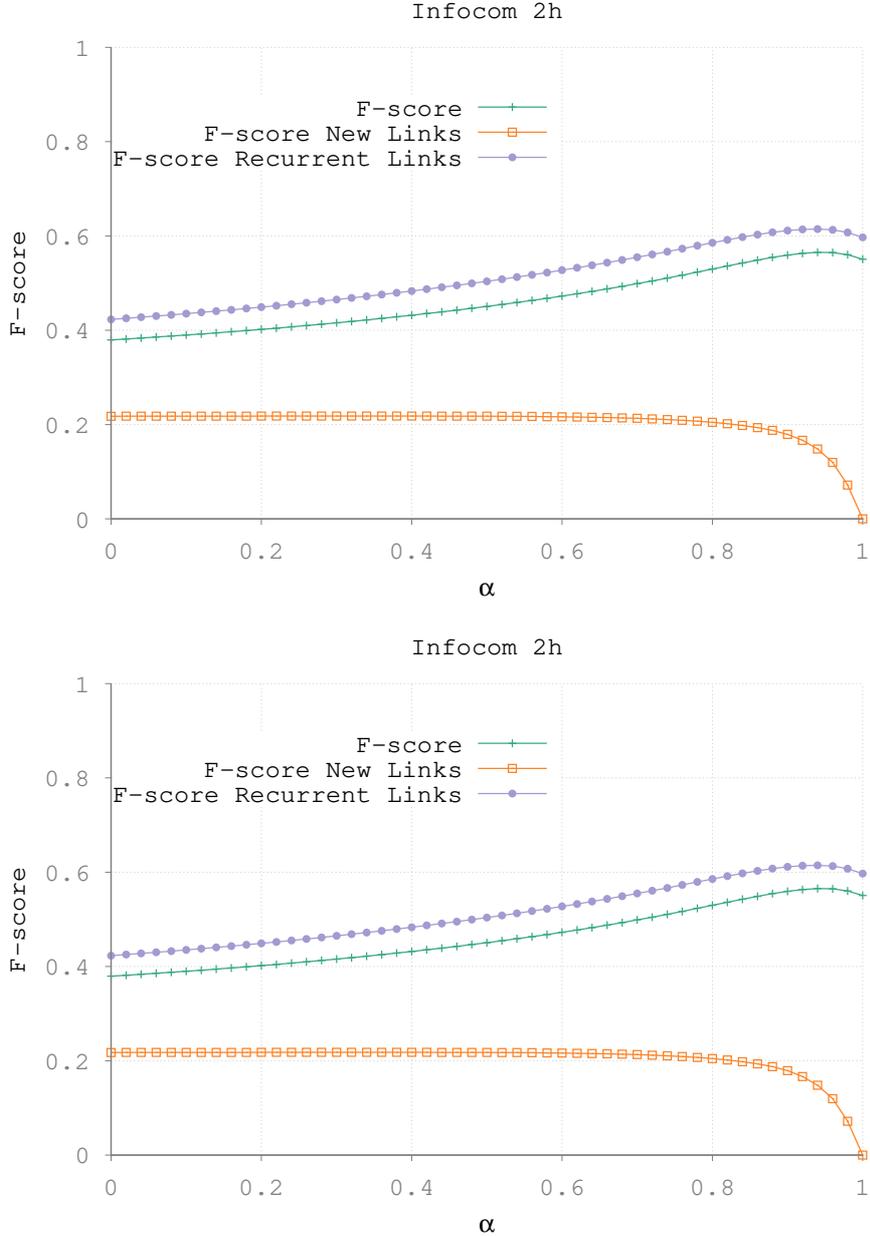}
				\includegraphics[width=12cm]{fullcommon.eps}
				\caption{F-score of the predictions for the mixing between \textit{PAE1000S}  and \textit{CN} as a function of $\alpha$ for different categories of node pairs for the \textit{Highschool} (top) and \textit{Infocom} (bottom) datasets using respectively 1 hour and 2 hours long observation and predictions periods.
					Green crosses: All links, Orange squares: New links, Purple circles: Recurrent links}
				\label{CNbench}
			\end{center}
		\end{figure}

			We display in Figure~\ref{CNbench} the obtained F-score as a function of $\alpha$ for the two categories of pairs aforementioned on both datasets as well as for the complete set of pairs.
			We can see on the \textit{Highschool} dataset that the F-score corresponding to the recurrent link category increases to a maximum for $\alpha \simeq 0.98$, while for the new link category it remains almost constant until $\alpha \simeq 0.98$, at which point it decreases to zero. 
			The F-score for the complete set of links is the same that previously, with the maximum at $\alpha=1$.

			\textit{Activity during the last 1000 seconds} alone is not able to predict new links and thus yields a null F-score. 

				The performance of the prediction of recurrent links improves as more weight is given to the \textit{Pair Activity Extrapolation}.
				However, we do not see the same effect of stagnation for a wide range of $\alpha$.

				The plot corresponding to the \textit{Infocom} dataset shows a quite different behavior.
				We can see that the F-score accounting for new link predictions starts from $0.22$ for $\alpha=0$ and slowly decreases until $\alpha=0.8$ at which point it sharply decreases to 0.
				Regarding the recurrent links prediction the F-score starts from $0.41$ and reaches a maximum of $0.62$ for $\alpha \simeq 0.92$ and then decreases to $0.56$.	

				We observe that new link prediction quality is noticeably lower than the recurrent link prediction quality in both experiment.
				The difficulty of this task is mainly due to the class imbalance problem, which is a well-known issue in the field of link prediction~\cite{lichtenwalter2010new}. 
				It is related to the fact that the number of links actually occurring is small in regards to the potential number of links, which is simply the number of node pairs.
				We observe that the number of links predicted in each category plays an important role in the prediction quality.

					Obviously, different metrics tend to predict preferentially different types of activity.
					These experiments show that, by choosing specific metrics combination, the prediction focuses on different kinds of activity, involving different kinds of links.
					With that in mind, we can put forward an explanation for the lower weights of structural metrics in the combinations computed by the learning algorithm in Section~\ref{sec:feat-comb}.
					Indeed, by only selecting temporal metrics the algorithm avoids the difficult task of predicting new links to favor the easier task of predicting recurrent ones.
					While it improves the overall F-score, it is also detrimental to the variety of links predicted.
					To counterbalance this unwanted behavior, we explore in the next section how to integrate the notion of category to the protocol.

				\section{Classes of pairs}
				\label{Classes}
				As seen in Section~\ref{Categories analysis}, specific metrics combinations favor different kinds of node pairs.
				However, our protocol does not make this distinction.
				Thus, we improve our approach by introducing classes of node pairs. 
				We compute a specific set of values for parameters $\alpha_m$ for each of these classes, allowing to adapt metric weights to each type of node pairs.

				\subsection{Class definition and metric to optimize}

				We choose to separate classes by their level of activity. 
				The underlying idea is to create classes reflecting on the one hand link prediction in graphs and in the other hand time series prediction.
				Indeed, predicting node pairs with no or low past activity is a task related to the prediction of new links, while predicting the future activity of recurrent links resembles a time series prediction task.

					We define three classes of node pairs. 
					Class C1 is populated with node pairs which have not interacted during the training period. 
					Class C2 gathers the pairs with less than a given number of links $k$ during the training period. 
					Lastly, the remaining high activity pairs are assigned to the class C3.
					Formally:
					\[  C1 = \{(uv) \in V\otimes V,~ |(t,uv) \in E_1|=0 \} \]
					\[  C2 = \{(uv) \in V\otimes V,~ 0 < |(t,uv) \in E_1| \leq k \} \]
					\[  C3 = \{(uv) \in V\otimes V,~ |(t,uv) \in E_1|=0 > k \} \]
					In the following experiments, in a {\em proof of concept} spirit we set the threshold between the class C2 and C3 to $k=5$ links.

					We follow the same experimental protocol for each class, using different values of parameters, which are computed during the learning phase.
					However, if we aim at optimizing the overall F-score, this protocol tends to favor high activity classes.
					Therefore, we have to update adequately the metrics to optimize with the gradient descent algorithm, in order to ensure we predict a wide variety of links.
				We then define the prediction score $\overline{F}$ as the harmonic mean of the F-scores for each class, that is 
				\[  \frac{1}{\overline{F}} =\frac{1}{3} \left ( \frac{1}{F(C_1)}+\frac{1}{F(C_2)}+\frac{1}{F(C_3)} \right )   \]
				where $F(C_1)$, $F(C_2)$ and $F(C_3)$ and the F-scores computed over each subset of node pairs defined by our classes.

			Once the parameters are optimized, we reassign each node pairs in our different classes based on their activity during the observation stream $L_2$. We then follow our protocol, combining the metrics for each classes using its respective set of parameters.

			\subsection{Reduction of the parameter space}
			
			In order to achieve computations in a reasonable amount of time, we reduce the set of metrics compared to the experiments in Section~\ref{sec:feat-comb}.
			In these series of experiments, we use the following metrics: \textit{number of Common Neighbors}, the \textit{Sorensen Index}, the \textit{Weighted number of Common Neighbors} and our temporal metrics, \textit{Pair Activity Extrapolation }(\textit{PAE}) , \textit{PAE10L}, the activity during the last 10 links, \textit{PAE1000S} and \textit{PAE10000S}, the activity during the last 1000 and 10000 seconds respectively, drawing benefit from the fact that several metrics are highly correlated.
			
\begin{figure}[h!]
				\begin{center}
					
					\includegraphics[width=9cm]{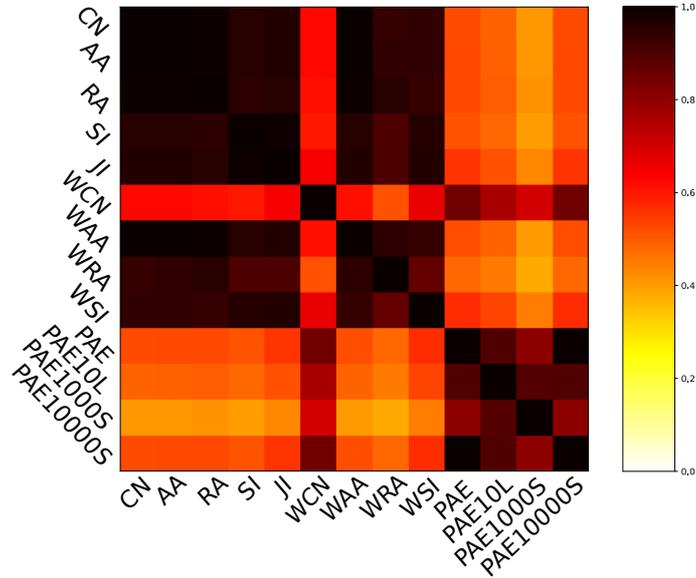}
					\caption{Correlation matrix of the values of each metric over all pair of nodes during the 2 hours training phase on \textit{Infocom}.}
					\label{coefmatrix}	
				\end{center}
			\end{figure}

			We can see in Figure~\ref{coefmatrix} the correlation matrix of the score given by each metric over every nodes for the training set of the experiment on \textit{Infocom} for periods of 2 hours.
			
			We can see that most structural metric are highly correlated. More particularly \textit{number of Common Neighbors}, \textit{Adamic-Adar index} and \textit{Resource Allocation index} seem to bring similar information. We also see that \textit{Sorensen Index} and \textit{Jaccard Index} are very similar. Therefore we choose one metric on each group, the \textit{number of Common Neighbors} and the \textit{Sorensen Index}. This will allow some variety in the structural metrics available.
			
			Similarly, except for the \textit{Weighed number of Common Neighbors}, the weighed variation of structural metrics are close to the other structural metric. Therefore, in the following experiments, we chose to keep the \textit{Weighted number of Common Neighbors}.
			The other temporal metrics tend to focus on different dynamics on the dataset and, while correlated, bring different information about the dynamic at different time scales in the system. Therefore we keep them for our following experiment.
			
		\subsection{Experimental results}

			We apply the protocol on \textit{Infocom} using the same training, validation, observation and prediction periods as in the 2 hours experiments detailed in Section~\ref{exp}.

			Results are summarized in Table \ref{Res Classes}. 
			We present the results for each class C1, C2 and C3, as well as the overall results obtained when combining the prediction for each class, denoted \textit{All Class}. 
			We also present the results obtained in the same experiments but without using classes of node pairs. 
			The corresponding results are prefixed C0. 	
			Thus, the F-scores associated with C0-1 C0-2 and C0-3 are the scores obtained when applying the evaluation protocol on the subset of node pairs relative to each pair classes.
		Precision, recall and F-score are reported, as well as overall activity predicted and occurring.
		These results correspond to the average over 10 runs of the experiments, with standard deviation below $0.01$ for the F-score, below $0.05 $ for the precision and recall, and below $22\%$ for the number of predicted links.

		\begin{center}
			\begin{table}[!h]
				
				\caption{F-score and number of link predicted and appeared by class on \textit{Infocom} using 2 hours periods.}
				\centering
				\begin{tabular}{| l | c | c | c | c | c | }
					\hline 
					Class & F-score  & Precision  & Recall  & Pred  & App  \\ \hline \hline
					C0 & 0.55 & 0.50 & 0.60  & 16737 & 14051 \\ \hline 
					C0-1 & 0.05 & 0.31  & 0.02  & 209  & 2668  \\ \hline 
					C0-2 & 0.37  & 0.46 & 0.30  & 1400  & 2134  \\ \hline 
					C0-3 & 0.63  & 0.51  & 0.84  & 15126  & 9137  \\ \hline \hline
					AllClass & 0.53  & 0.49  & 0.59  & 16737  & 14051  \\ \hline 
					C1 & 0.25  & 0.23  & 0.29  & 3300  & 2668 \\ \hline 
					C2 & 0.41 & 0.37  & 0.48 & 2800  & 2134 \\ \hline 
					C3 & 0.65 & 0.60  & 0.70  & 10500  & 9137 \\ \hline 
				\end{tabular}
				\label{Res Classes}
			\end{table}		
		\end{center} 
		
		First, we observe that the number of predicted links in class C1 is more than 15 times higher than in class C0-1.
		The precision slightly decreases, but the recall is so much higher that the new protocol achieve a $0.25$ F-score, to be compared to the $0.05$ F-score obtained on this class in the standard series of experiments.
		We also note that the activity predicted over the whole class C2 nearly doubles, which also leads to a performance improvement, from $0.37$ to $0.41$ F-score.
		It is very interesting to note that the introduction of classes also improves the prediction in the classes C3, from a $0.63$ to a $0.65$ F-score.
		The number of links predicted in this class largely decreases to balance the fact that the activity predicted in the two other classes increases, making it closer to the actual activity observed.
		We investigate this effect in more details later.

		Quite counter-intuitively, while prediction is improved in each class, the F-score predicted overall decreases from $0.55$ to $0.53$.	
		This phenomenon is a sort of amalgamation paradox.
		Precisely, the F-score is the harmonic mean of the precision and the recall, hence its computation is related to the proportion of links that should be predicted and that is actually predicted in each class. 
		This may lead to the fact that the F-score over the union of subsets and the F-scores of each subset considered separately evolve according to different trends.

\begin{figure}[h]
			\begin{center}
				
				\includegraphics[width=12cm]{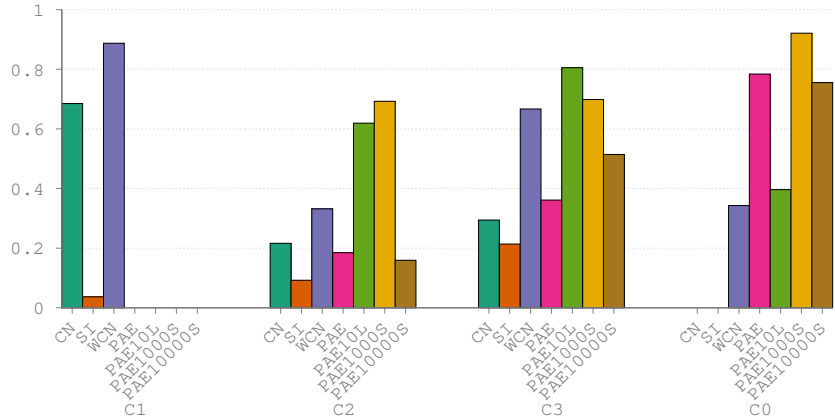}			
				
				\caption{Metric coefficients repartition computed by the learning algorithm during prediction by classes on \textit{Infocom} using 2 hours periods.}
				\label{Bars Classes}
			\end{center}
		\end{figure}

			We display in Figure~\ref{Bars Classes}, the values of parameters $ \alpha_m$ computed by the learning algorithm to achieve a prediction over each class (averaged over 10 experiments).
			As expected, the method tends to use different combinations to make a prediction over each class. 
			Considering class C1, the algorithm favors structural and hybrid metrics. 
			These metrics are commonly used for link prediction in graphs, which makes sense given that predicting new link activity is closely related to this problem.

			Class C2 tends to mix structural metrics to temporal ones.
			Pairs in this class have intermediary behavior and the algorithm combines information of different nature to predict these behaviors. 
			Interestingly, prediction on class C3 also mixes temporal metrics to structural ones, even if temporal metrics remain predominant.	
			It contrast with prediction in the experiments without class, where no or little weight was distributed to purely structural metrics, and little weight to hybrid ones.
			Our interpretation is that in the former protocol, the gradient descent favored temporal metrics in order to exclude node pairs which had never interacted and then are much more risky to predict, while in the later protocol, the introduction of classes allows to focus on recurrent links only, so that giving weight to structural metrics becomes interesting an option again.

		Experiments on \textit{Highschool} and \textit{Reality Mining} presented similar results and are not presented here. Concerning the \textit{Taxi} dataset, we think that the different nature of interactions between nodes necessitate a specific analysis which is left for future works. 
		
		In the light of these experiments, we can see that introducing classes of pairs in the problem of activity prediction may be a good option in order to improve the diversity of links to be predicted, which opens interesting leads for future works on the topic.

		\section{Conclusion}

			In this work, we proposed an activity prediction protocol adapted to the link stream formalism, making it possible to advantageously use the rich information contained in this modeling. 
			It is relies on a flexible way to combine the information from metrics which capture metrics of the stream.
			We also proposed an evaluation protocol adapted to our problem. 
			Our experiments show that our protocol is able to find efficient combinations of structural and temporal metrics that lead to performance improvements, compared to benchmarks such as the past activity extrapolation.
			We also investigated how this algorithm tends to exclude specific types of node pairs, leading to less variety in the predicted links.
			However, we showed that it is possible to mitigate this issue by introducing classes of node pairs in activity prediction, so that a better balance of activity can be obtained in each class.
			Our protocol is designed in a modular way, such that each part is independent from the others and can be replaced or improved, depending on the application we are interested in.

			Different improvements are considered for future works. 
			The metrics presented in this work are classical metrics used for link prediction in graphs or basic ways to capture the temporal information of the stream. 
			As our protocol is ready to combine new metrics, we intend to design refined ones that are able to detect more subtle dynamical metrics of the stream, for example implementing pattern mining techniques to identify typical motifs of the short term dynamics. 
			We also made the assumption that the activity remains constant from the observation period to the prediction period. 
			However, this hypothesis is not always satisfied and greatly depends on the data under concern. 
			Models developed in the context of time series prediction, like the ARIMA model which extrapolates precisely past activity~\cite{huang2009time}, would certainly allow to better evaluate the number of predicted links.
			Finally, we also want to investigate further how our algorithm behaves with different classes of pairs. 
			Other definitions are possible, for example structural classes, where node pairs of a same community would belong to a same class.
\subsection*{Acknowledgements}
This work is supported in part by the European Commission H2020 FETPROACT 2016-2017 program under grant 732942 (ODYCCEUS), by the ANR (French National Agency of Research) under grants ANR-15-CE38-0001 (AlgoDiv) and by the Paris Ile-de-France Region and program FUI21 under grant 16010629 (iTRAC).

\bibliographystyle{abbrv}
\bibliography{MyCollection}   


\end{document}